# Evaluation of polarization characteristics in metal/ferroelectric/semiconductor capacitors and ferroelectric field-effect transistors

Kasidit Toprasertpong,[a)] Kento Tahara, Mitsuru Takenaka, and Shinichi Takagi

*Department of Electrical Engineering and Information Systems, The University of Tokyo, 7-3-1 Hongo, Bunkyo-ku, Tokyo 113-8656, Japan*

In this study, we propose a measurement technique for evaluating ferroelectric polarization characteristics in ferroelectric field-effect transistors (FeFETs). Different from standard metal/ferroelectric/metal capacitors, the depletion and inversion phenomena in semiconductor substrates have to be carefully taken into account when evaluating the ferroelectric properties using fast voltage sweep as input. The non-equilibrium deep depletion is found to be the limiting factor for the accurate evaluation of ferroelectric properties in metal/ferroelectric/semiconductor capacitors. By connecting the source, drain, and substrate of the FeFET together during the polarization measurement, the deep depletion can be suppressed and the ferroelectricity of the ferroelectric gate can be accurately evaluated. The present technique is a powerful method for capturing the polarization states in FeFETs, enabling new approaches for device characterization and fundamental study, and overcomes the limitation found in the conventional polarization measurement on 2-terminal metal/ferroelectric/semiconductor capacitors.

Ferroelectric field-effect transistors (FeFETs), in which ferroelectric thin films are employed as FET gate insulators, are promising for non-volatile memory devices[1-3] and synaptic devices in computing-in-memory applications.[4-6] A substantially increasing interest in FeFETs can be seen after the discovery of ferroelectric $HfO_2$[7] thanks to its scalability and CMOS compatibility,[3] with FeFETs having excellent performance being continuously reported.[8-11] So far, the characterization of FeFETs is mainly limited to the standard evaluation of $I_d$-$V_g$ characteristics and threshold voltage shift $\Delta V_T$ as well as their retention and endurance properties.[3,5,8-17] On the other hand, the ferroelectric properties of the ferroelectric gate insulators are usually evaluated in an indirect manner, by preparing separate metal/ferroelectric/metal (MFM) capacitors for characterization.[3,5,8,14-16] It is usually assumed that the ferroelectric properties of the separately-prepared MFM capacitors can well reproduce those of the ferroelectric gate insulators in the FeFETs. However, it is questionable whether MFM capacitors can represent the ferroelectricity of the ferroelectric gate insulators, particularly for metal/ferroelectric/semiconductor (MFS, or sometimes called MFIS if the dielectric insulator layer should be emphasized)-type FeFETs, as it is well recognized that the ferroelectric properties of $HfO_2$ is strongly dependent on the underlying and capping layers.[17-20]

To carry out more accurate characterization, MFS capacitors are frequently employed to reproduce the same substrate condition as the ferroelectric gates in FeFETs.[17,21,22] By characterizing MFS capacitors, it has been reported that semiconductor substrates have a significant effect on the ferroelectric properties.[23,24] On the other hand, there are two concerns with the characterization of separate MFS capacitors as a method to predict the ferroelectricity of ferroelectric gates insulators in FeFETs. First, high doping concentration in the order of $10^{19}$ to $10^{20}$ cm$^{-3}$ is usually employed in the semiconductor substrates of MFS capacitors (called here as MFS$_+$) to minimize the substrate depletion. Due to different band bending in different substrate doping concentrations, the voltage across the ferroelectric layer in MFS$_+$ capacitors cannot reproduce the actual voltage across the ferroelectric layer in FeFETs, making it difficult to be used to predict the polarization-voltage (P-V) hysteresis loops in actual FeFETs. The other concern is that the behavior of ferroelectric properties found only in FeFETs, such as influence originating from thermal or plasma processes during FeFET fabrication and the FET geometry, cannot be investigated by simply characterizing MFS capacitors. In this study, we propose a measurement technique to evaluate the P-V hysteresis characteristics of the ferroelectric gate insulators directly from the FeFET structure. We address how the P-V hysteresis measurement directly on FeFETs is different from the conventional P-V measurement on MFS capacitors, where the deep depletion phenomenon plays a substantial role during measurements.

To study the behavior of polarization characteristics obtained from P-V measurements on capacitors, ferroelectric capacitors consisting of ferroelectric 10-nm-thick $Hf_{0.5}Zr_{0.5}O_2$ (HZO) were prepared in MFM, MFS$_+$, and MFS$_s$ (standard MFS) configurations. For the MFM capacitor, 13-nm-thick TiN was sputtered on a heavily-doped p-Si substrate ($N_A \approx 10^{20}$ cm$^{-3}$) before HZO was deposited by an atomic layer deposition process. The device was subsequently capped by 13-nm-thick TiN to form an MFM structure and annealed at 400°C for 30 s in a nitrogen atmosphere to crystalize the ferroelectric phase. For the MFS$_+$ capacitor, the fabrication process was similar to the MFM capacitor except that the bottom TiN was removed, and for the MFS$_s$ capacitor, the heavily-doped p-Si substrate was further replaced by a p-Si substrate with a lower doping concentration ($N_A \approx 10^{15}$ cm$^{-3}$), which is also used later in this study for fabricating an FeFET. Except the substrate preparation before the HZO deposition, the fabrications of the MFM, MFS$_+$, and MFS$_s$ capacitors were carried out simultaneously in the same process line for a fair comparison.

Figure 1(a) shows the schematic of a conventional measurement setup for evaluating P-V hysteresis characteristics in ferroelectric capacitors.[25] Triangular-wave voltage, typically with a frequency of the order of 1 kHz to 1 MHz, is applied to one of the metal plates of the capacitor.

[a)]toprasertpong@mosfet.t.u-tokyo.ac.jp

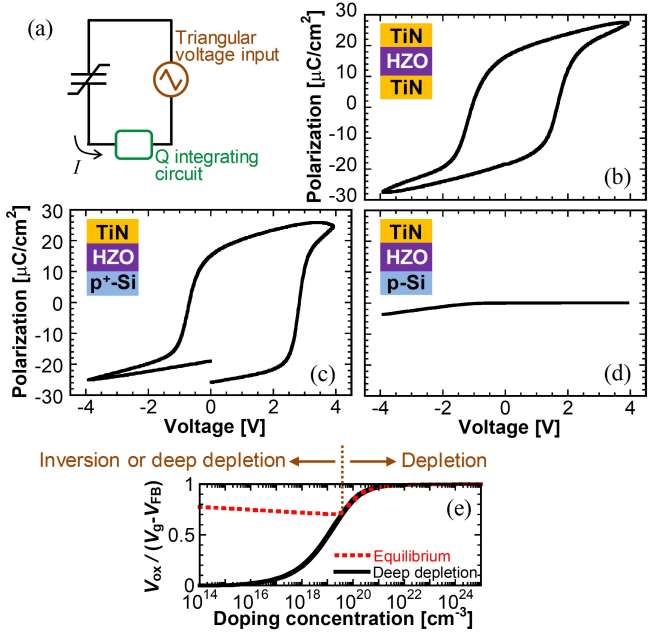

**FIG. 1.** (a) Schematic of *P-V* hysteresis measurement on ferroelectric capacitors. Measured *P-V* characteristics of (b) TiN/HZO/TiN MFM, (c) TiN/HZO/p$^+$-Si ($N_A \approx 10^{20}$ cm$^{-3}$) MFS$_+$, and (d) TiN/HZO/p-Si ($N_A \approx 10^{15}$ cm$^{-3}$) MFS$_s$ capacitors at 1 kHz. (e) Calculated ratio of voltage $V_{ox}$ across the gate insulator to the applied voltage $V_g - V_{FB}$ in MFS capacitors with given Si substrate doping concentrations assuming $V_g - V_{FB} = 4$ V, the equivalent oxide thickness of 2.6 nm, and both the equilibrium and deep depletion conditions.

The polarization switching during the voltage sweep is evaluated as the amount of charges leaving the other side of the capacitor. The *P-V* characteristics of the fabricated capacitors using 1-kHz triangular-voltage input are shown in Figs. 1(b)-(d). The MFM capacitor shows a typical polarization hysteresis loop [Fig. 1(b)] with $2P_r = 31.1$ μC/cm$^2$ ($2P_r$ values in this paper were evaluated from *P-V* characteristics after leakage compensation[26] as shown in Fig. 1S in supplementary material) and a clear polarization switching behavior. The MFS$_+$ capacitor show a polarization hysteresis loop similarly to the MFM capacitor [Fig. 1(c)] with slightly larger $2P_r = 33.8$ μC/cm$^2$. The slight difference in the $2P_r$ values between the MFM and MFS$_+$ capacitors is possibly due to the difference in the ferroelectric phase crystallization caused by different boundary conditions of HZO during annealing, implying that the MFM characteristics are not an accurate measure of the ferroelectric properties in MFS-type FeFET ferroelectric gates.

Different from simple MFM capacitors, the applied voltage to MFS capacitors appears not only across the ferroelectric layers but also partly across the semiconductor substrates[27] (and dielectric layers, if any, but will not be discussed in this letter for simplicity). Figure 1(e) shows the fraction of applied voltage divided to the ferroelectric insulator of MFS capacitors with substrate doping concentration $N_A$. The polarization switching is not taken into account in the calculation for simplicity. It can be seen that $N_A$ as high as $10^{20}$ cm$^{-3}$ still shows a finite voltage loss on the substrate and cannot be simply assumed as a metal-like substrate. Moreover, as can also be seen in Fig. 1(e) that the voltage $V_{ox}$ across the ferroelectric layer depends on the substrate doping concentration, it is not straightforward to compare MFS$_+$ capacitors having heavily-doped substrates with FeFETs whose substrates have much lower doping concentration. For substrates with low $N_A$, the formation of an inversion layer plays an important role in determining electric field distribution. The inversion layer is slightly more difficult to be formed in a substrate with higher $N_A$, resulting in a larger voltage loss [~$2\phi_B$ surface potential ∝ ln($N_A$)] on the substrate as can be seen in a small decrease of $V_{ox}$ in the mid-range of $N_A$ in Fig. 1(e). On the other hand, for considerably high $N_A$, there is no inversion layer formed. The space charge in a small band bending (<< $2\phi_B$) is sufficiently large to realize the total capacitor charge, resulting in a small voltage loss on the substrate and larger $V_{ox}$. Hence, it is clear that the analysis of voltage loss on the semiconductor substrate strongly depends on the substrate doping concentration $N_A$.

To evaluate the ferroelectric properties of the gate insulators in FeFETs in a more accurate manner, one may expect the *P-V* characterization to be performed on MFS$_s$ capacitors, whose substrates are the same as ones of FeFETs. However, the *P-V* result of the MFS$_s$ capacitor, shown in Fig. 1(d), exhibits non-hysteretic characteristics and the ferroelectric properties cannot be observed, even though HZO is confirmed to be ferroelectric from the ferroelectric properties in the MFS$_+$ capacitor. Thus, the peculiar *P-V* characteristics observed in Fig. 1(d) is attributed to the deep depletion phenomenon well recognized in typical Si metal/oxide/semiconductor (MOS) capacitors. When the sweep rate of voltages applied on MOS capacitors is so fast that the equilibrium condition cannot be maintained, the inversion layers cannot be formed and deep depletion occurs. The critical voltage-sweep rate to maintain equilibrium given by[28]

$$(dV/dt)_c = q n_i W / \tau_g C_{ox}, \quad (1)$$

where $q$ is the elementary charge, $n_i$ the intrinsic carrier concentration, $W$ the depletion width, $\tau_g$ the carrier generation time, and $C_{ox}$ the oxide capacitance, is estimated to be in the order of 0.1–1 V/s for the MFS$_s$ capacitor in this study ($n_i \approx 10^{10}$ cm$^{-3}$, $W \approx 1$ μm, $\tau_g \approx 0.1$ μs[29], $C_{ox} \approx 1$ μF/cm$^2$). On the other hand, *P-V* measurements typically employ high voltage-sweep rates more than 1 kV/s to suppress the influence of leakage current, which is superimposed on the measured *P-V* characteristics by a factor of $1/(dV/dt)$;[30] for instance, the 1-kHz *P-V* measurements in this study used a sweep rate of 16 kV/s. The large difference between the voltage-sweep rate required to maintain equilibrium and sweep rates used for *P-V* measurements results in the non-equilibrium condition occurring in MFS capacitors during *P-V* measurements.

The MFS$_s$ bands diagram during *P-V* measurements is illustrated in Fig. 2(a). When deep depletion occurs in substrates with doping concentration $N_A$ less than $10^{19}$ cm$^{-3}$ [Fig. 1(e)], most of the applied voltage will appear across the substrate and the electric field across the ferroelectric layer becomes much smaller than the coercive field $E_c$ needed for the ferroelectric polarization reversal [Fig. 2(b)]. In this situation, even though the polarization can be switched once when the MFS$_s$ capacitor is under the accumulation condition, it cannot be switched back under the deep depletion condition. As a result, the polarization is kept fixed in one state after the first measurement and the MFS$_s$ capacitor thereafter behaves as a MOS capacitor with no detectable polarization switching. Figure 2(c) shows the *C-V* characteristics of the MFS$_s$ capacitor at a small-signal frequency of 100 kHz. The *C-V* characteristics similar to those in typical non-ferroelectric MOS capacitors with electron-trap hysteresis supports our above explanation. Note that the *C-V* measurement, using a voltage sweep rate of 0.4 V/s much slower than in *P-V* measurements, shows a clear deep depletion behaviour, where capacitances reach lower values than the minimum equilibrium high-frequency capacitance $C_{min}$. Interestingly, the peculiar *P-V* result, a line with an almost constant slope at negative voltages and a flat line at positive voltages, can be precisely described by the integration of the *C-V* curve as shown in Fig. 2(d) since both measurements similarly describe the total charges on capacitors. Note that this relation between *P-V* and *C-V* results does not hold if the ferroelectric polarization can be properly switched: *P-V* measurements reflect total polarizations whereas *C-V* measurements reflect only polarizations that response to the small-signal input.[31,32] A good match of the *P-V* and integrated *C-V* curves is thus another supportive evidence that the polarization cannot be properly switched and the ferroelectricity cannot be evaluated in the MFS$_s$ capacitor exhibiting deep depletion.

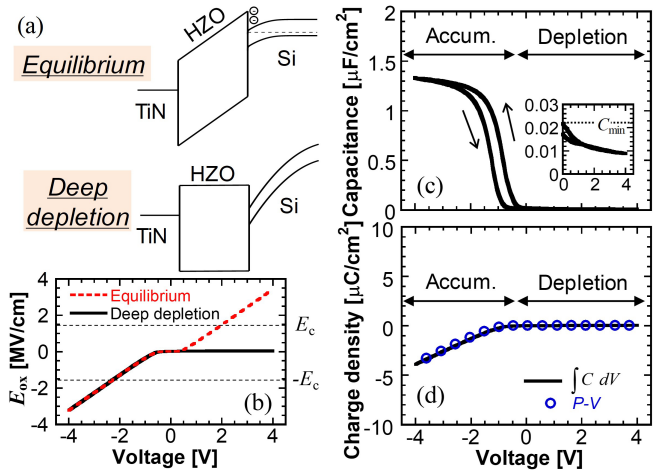

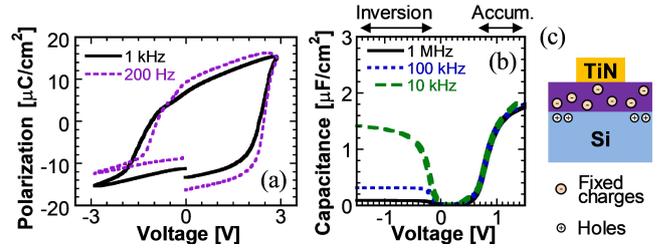

FIG. 2. (a) Band diagrams and (b) average oxide fields $E_{ox}$ of MFS$_s$ capacitors at the equilibrium and deep depletion conditions when $V_g$ = 4 V is applied. The thicknesses of Si bands are scaled with a 1:100 ratio. The interfacial layer is neglected for simplicity. (c) *C-V* characteristics of the MFS$_s$ capacitor at 100 kHz. $C_{min}$ in the enlarged inset indicates the minimum capacitance expected when the strong inversion occurs at the equilibrium. (d) Charge density accumulated in the MFS$_s$ capacitor obtained from *C-V* characteristics. Open circles correspond to the *P-V* characteristics in Fig. 1(d).

It should be addressed that *P-V* measurements on MFS capacitors with low substrate doping concentration may provide ferroelectric hysteresis characteristics in some conditions that inversion carriers can quickly reach equilibrium, for example in substrates with narrow bandgaps. Furthermore, as shown in Fig. 3(a), the polarization hysteresis characteristics can be obtained from a *P-V* measurement on an MFS$_s$ capacitor prepared in the same way but on an n-type Si substrate with $N_D \approx 10^{15}$ cm$^{-3}$, suggesting that inversion holes can be formed to suppress deep depletion. This is attributed to negative fixed charges in the HZO film. The *C-V* characteristics of the MFS$_s$ capacitor on the n-Si substrate in Fig. 3(b) shows such a large flat-band shift that the inversion condition occurs at $V$ = 0 V, implying that there are a large amount of negative fixed charges in the HZO film.[33] These negative fixed charges induce inversion holes in the peripheral regions surrounding the gate electrode of the capacitor, which in turn become a reservoir of holes to assist the inversion hole formation beneath the gate electrode [Fig. 3(c)] and help suppress voltage drop across substrates during *P-V* measurements [Fig. 1(e)]. The difference in the *P-V* results only by changing substrate types in this study indicates that we should be careful when analyzing the *P-V* characteristics of MFS capacitors: it is significantly affected by the formation capability of inversion layers and does not accurately reflect the ferroelectric properties. Note that fixed charges are sensitive to the fabrication process and, thus, the same tendency might not be obtained in devices with different process conditions.

FIG. 3. (a) *P-V* and (b) *C-V* characteristics of the MFS$_s$ capacitor prepared on the n-Si substrate. Inversion charge response can be observed. (c) Charge distribution in the MFS capacitors prepared in this study. Negative fixed charges in the HZO layer induce holes surrounding the MFS capacitors. The supply of holes from the surrounding helps suppress the deep depletion in the n-Si substrate and assures sufficient electric field in the ferroelectric layer for the polarization reversal.

Even though *P-V* measurements might be possible under the inversion of the surrounding substrate surface, the inversion layer formation is not necessarily in equilibrium and thus measurement in this way is not an effective approach to obtain accurate polarization characteristics. As a solution to the deep depletion issue during fast-sweep-rate *P-V* measurements, we propose a *P-V* measurement technique to evaluate the ferroelectric properties by employing FeFETs as illustrated in Fig. 4(a). By additionally connecting the substrate terminal to the source and drain (S/D) contacts, the majority carriers can be supplied through the substrate contact and the minority carriers through the S/D contacts,[32] assuring both carriers in the substrate to be in equilibrium. As a demonstration, we prepared an FeFET using the same p-Si substrate and the same fabrication line for the gate stack as those of the MFS$_s$ capacitor in Fig. 1(d). As shown in Fig. 4(b), the FeFET shows $I_d$-$V_g$ characteristics (drain voltage $V_d$ = 100 mV) with ferroelectric hysteresis, indicating that the fabricated device operates properly as a so-called FeFET. By connecting the substrate and S/D contacts together, a clear $P$-$V$ hysteresis loop [Fig. 4(c)] as well as displacement current with polarization switching peaks [Fig. 4(d)] can be obtained. This allows us to accurately evaluate the ferroelectric properties of the ferroelectric material stacked on semiconductor substrates. As a result, $2P_r$ = 43.5 μC/cm$^2$ is obtained from the fabricated FeFET. $2P_r$ larger than that measured on the MFS$_+$ capacitor is possibly owing to different charge distribution at the semiconductor/ferroelectric interfaces. Whereas the substrate of the MFS$_+$ capacitor is not inverted, it has been reported that the inversion layer in FeFETs is trapped near the semiconductor/ferroelectric interface which in turn increases the electric field across the ferroelectric layer.[32] As shown in the inset of Fig. 4(c), the result of the $P$-$V$ measurement on the FeFET when the S/D is not connected with the substrate and kept floating is similar to the $P$-$V$ result of the MFS$_s$ capacitor, meaning that electrical connection of the S/D is necessary to assure the equilibrium state of the MFS stack.

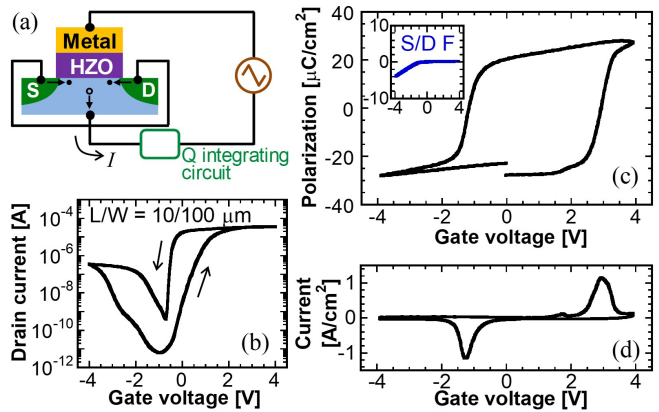

FIG. 4. (a) Schematic of *P-V* hysteresis measurement on FeFET. The S/D and substrate of the FeFET are connected together to measure the total charges induced at the semiconductor side. (b) $I_d$-$V_g$ characteristics of the fabricated FeFET at $V_d$ = 100 mV. (c) *P-V* characteristics and (d) displacement current of the FeFET measured at 1 kHz when the S/D and substrate are connected. The inset is the *P-V* characteristics of the same FeFET when the S/D are not connected to the *P-V* setup and kept floating.

The proposed technique is a useful method to characterize the ferroelectric properties of the FeFET gate stack directly from FeFETs. The evaluated ferroelectric properties will include any possible degradation or enhancement of the ferroelectric phase caused by the FET fabrication process as well as any possible back-end-of-line processes. Furthermore, the $P$-$V$ characteristics in this way, which can be called as the $P$-$V_g$ characteristics, enable us to capture the polarization state at a given applied gate voltage. Together with the $I_d$-$V_g$ characteristics, the accurate polarization states help us get insights into the device operation of FeFETs, which has not been fully understood yet. This includes analyses of negative-capacitance FeFETs (NCFET),[34] promising low-power FET devices based on a controversial assumption of the S-curve polarization properties. A controversy on the device operation of NCFETs[35-37] is partly due to a lack of techniques to directly acquire the $P$-$V_g$ characteristics of NCFET devices, limiting the discussion of the impact of the NC effect on the subthreshold swing.[34-37] Using the $P$-$V_g$ technique in this study allows us to distinguish ferroelectric-gate devices under test whether they are operating as memory FeFETs with polarization switching or NCFETs with stable non-hysteresis polarization characteristics.

In summary, we have proposed a $P$-$V_g$ measurement technique to directly evaluate the ferroelectric properties in the ferroelectric gate stacks of FeFETs without having to prepare additional test capacitors. The importance of maintaining equilibrium for both electrons and holes in the semiconductor substrates on accurate evaluation of polarization characteristics has been pointed out by carrying out the systematic comparison with conventional $P$-$V$ measurements on MFS capacitors. This technique allows us to capture the polarization state in the ferroelectric gates at given device conditions, providing key information toward characterization of FeFETs, such as $2P_r$ values and their retention and endurance properties, as well as toward comprehensive understanding of underlying device physics of FeFETs.

See the supplementary material for the $P$-$V$ characteristics after the leakage compensation.

Appl. Phys. Lett. **116**, 242903 (2020); https://doi.org/10.1063/5.0008060

This work was supported by JSPS KAKENHI Grant No. 19K15021, 17H06148 and JST-CREST Grant No. JPMJCR1332, Japan.